# ACO Implementation for Sequence Alignment with Genetic Algorithms

Aaron Lee                    Livia King

**Abstract**

In this paper, we implement Ant Colony Optimization (ACO) for sequence alignment. ACO is a meta-heuristic recently developed for nearest neighbor approximations in large, NP-hard search spaces. Here we use a genetic algorithm approach to evolve the best parameters for an ACO designed to align two sequences. We then used the best parameters found to interpolate approximate optimal parameters for a given string length within a range. The basis of our comparison is the alignment given by the Needleman-Wunsch algorithm. We found that ACO can indeed be applied to sequence alignment. While it is computationally expensive compared to other equivalent algorithms, it is a promising algorithm that can be readily applied to a variety of other biological problems.

## Motivation

Ant colony optimization (ACO) is a novel algorithm recently developed for nearest neighbor approximations in large, NP-hard search spaces [1]. The algorithm itself is more accurately described as a meta-heuristic that must be applied differently to different problems. ACOs have been applied successfully to both the Traveling Salesman Problem (TSP) [2] and the Quadratic Assignment Problem (QAP) [3].

More recently, better methods of attaining optimal values for ACOs have been developed. In 1998, Bottee et al. used genetic algorithms on top of ACOs to achieve locally optimal values for ACOs very quickly [4].

As far as we have been able to tell, the ACO algorithm has not yet been applied to biological problems. Most biological problems today face the same difficulty as other NP-hard computational problems. The ACO algorithm seems particularly well-suited for these kinds of applications.

We propose using ant colony optimization with genetic algorithms and applying the resulting ACO to two sequence alignment.

## Introduction

Ant Colony Optimization is a meta-heuristic that was developed after studying the behavior of ants and learning how ants were able to solve constraint optimization problems in nature. It was found that ants laid down pheromone trails as they traversed the ground. As soon as a food source was found, the ants would start to wander back along the path that they had traversed to the ant colony. Because the ant who came back first with food would double the pheromone on the trail, the next ant to leave the ant colony in search of food who have a higher chance to take that path because ants will tend to follow trails that have the highest pheromone concentration [1].

In nature, it was found that this often led to suboptimal solutions. The decay of the pheromone was occasionally not fast enough to cause the ants to seek other solutions; hence, there was a slight propensity to follow the path of the ant that returned back to the ant colony first. In other words, there was too much of a weight on the ant that reached the first food source, because the first food source found was not necessarily the closest food source globally.

In computers, this limitation was easily avoided by adding a pheromone evaporation effect where the ants' pheromones would decrease in concentration as time progressed, preventing the early convergence onto local minimums [5].

## Critical Review

Dorigo, Caro, and Gambardella outlined in their 1999 paper the basic characteristics of the ant algorithm and reported on some of the uses for which it has been tested. They list the ant algorithm as deriving many of its characteristics from real ant colonies. For example, artificial ants, like real ants, are part of a colony of cooperating individuals, and use pheromones to facilitate indirect communication between themselves and other ants. Although each of the individuals in a population is simple and not capable of deriving a good solution to a problem, the population as a whole can come up with a good solution, concept that is now known as swarm intelligence. Both real ants and artificial ants make decisions using local information.

Unlike real ants, however, artificial ants have some abilities that allow them to derive higher quality solutions than their natural counterparts. Artificial ants can keep a memory of past actions. Their methods of pheromone depositing are also different. They may deposit varying amounts of pheromones based on the quality of the solution found and might deposit pheromones after the generation of a complete solution rather than step by step. Other abilities like the ability to look ahead several steps, or to backtrack, can also be added to artificial ants.

The first problem ant algorithms were applied to was the traveling salesman problem (TSP), a commonly researched NP hard problem in which a theoretical salesman has to find the shortest path connecting n cities. Three algorithms based on the ACO meta-heuristic were defined, and the "ant-cycle" algorithm (later called Ant System, or AS), a variation in which the ants deposited pheromone after they finished constructing their solution was found to be the most effective. The ant algorithm performed as well as several other general purpose heuristics on relatively small problems (involving 30 to 75 cities). However, the ant algorithm was not able to come up with the best known solution when it was applied to larger problems, although it was able to find good solutions quickly. Later variations on the AS algorithm, most notably ant colony systems (ACS), were more successful and beat many other heuristics in terms of solution quality and CPU time when run on standard TSP problems.

ACO was next applied to the Quadratic Assignment problem, which involves assigning $n$ facilities to $n$ locations so as to minimize cost. In this case, the ACO algorithms tested are the best available heuristic. ACO has also been shown to be the best available heuristic by far in the sequential ordering problem. Other problems to which ACO has been applied include Job shop scheduling, vehicle routing, graph coloring, shortest common supersequence, connection oriented network routing, and connection-less network routing.

From the results listed by Dorigo et. al, it can be seen that ACO is a valid approach that has been successful for many problems. However, there are some limitations to ACO's potential. First, because of the ant algorithm's random nature, the success of ant algorithms very problem dependent and it is hard to predict what variations on the algorithm will be successful without actually implementing the algorithm and testing it. There are also limitations as to which problems can be solved with ACO. The problem must lend itself to being represented by ants traveling across a virtual space. Each state available to the ant must also not have many neighbor states, as too many neighbor states will minimize the chance of two ants appearing in the same state, and render the pheromone communication worthless.

# Design

## ACO

We closely modeled our approach after Bottee and Bonabeau's algorithm for solving the traveling salesman problem.

A sequence alignment between two sequences n and m bases long is represented by a n x m Needleman-wunsch scoring matrix. At the beginning of every generation, a set number of ants start at the bottom right corner (which represents the end of an alignment) of the matrix, and traverse the matrix until they reach the beginning of the alignment (when the ant reaches either the top or left edge of the matrix). From any given position, an ant can move in one of three directions: up, left, or diagonally up-left. As is the case with the Needleman-Wunsch algorithm, a move in the former two directions represents a gap in one of the strands, while a diagonal move is either a match or a mismatch.

At every step, the ants use a scoring algorithm to decide what direction to move in. Every direction (up, left, or diagonal) is assigned a score based on the following formula:

$$\phi_i^{W_\phi} \cdot M^{W_m} \cdot R^{W_r}$$

Where $\phi_i$ is the pheromones level for step i in a particular direction, $M$ has a value of 2 if going in that direction will result in a match and 1 if it results in a mismatch, $R$ is the regional weighting score that encourages ants to move toward the diagonal of the matrix, and $W_\phi$, $W_m$, and $W_r$ are the weights assigned to those values. The ant chooses a number $p_i$. If $p_i$ is greater than the parameter $p$, then the ant simply moves in the direction with the highest score. If $p_i$ is less than $p$, the ant chooses the direction using a weighted probability based on the scores.

After each ant moves, it leaves a pheromone trail according to the formula.

$$\phi_{i+1} = (\phi_i + \phi_{step}) \cdot q_l$$

Where $\phi_i$ is the old pheromone level, $phi_{i+1}$ is the new pheromone level, and $q_l$ is a number between 0 and 1 that represents the local decay of pheromones.

As each ant moves, it also scores its path using a Needleman-Wunsch like scoring scheme that adds points for matches and subtracts points for mismatches and gaps. At the end of each generation, the best path is reinforced according to the formula.

$$\phi_{i+1} = (\phi_i + \phi_{step} \cdot (S_g/S_h)$$

| Variable | Range |
|----------|-------|
| $g$ | 10 - 40 |
| $a$ | 5 - 30 |
| $\phi_0$ | 0 - 1 |
| $\phi_{step}$ | 0 - 1 |
| $\overline{W}_\phi$ | 0 - 10 |
| $\overline{W}_m$ | 0 - 10 |
| $\overline{W}_r$ | 0 - 10 |
| $q_l$ | 0 - 1 |
| $q_g$ | 0 - 1 |
| $p$ | 0 - 1 |

**Table 1**: Variables optimized for in the genetic algorithm.

where $S_g$ is the score of the best path of this generation, and $S_b$ is the best score obtained so far overall. The entire matrix also undergoes a global decay:

$$\phi_{i+1} = \phi_i \cdot q_g$$

where $q_g$ is the global decay parameter.

The simulation keeps on looping through generations until the score for the best alignment converges, or until the maximum number of generations is reached.

## Genetic Algorithm

The motivation for using a genetic algorithm is simple: all the parameters above can be approximately determined using a genetic algorithm which evolves for the best set of parameters. Since hand-tweaking of ACO parameters can be a futile exercise and often a waste of time, genetic algorithms provide a convenient alternative albeit it may sometimes be inaccurate. We used a very simple implementation of genetic algorithms to quickly find a set of parameters that were locally optimal at discrete steps of search string length.

At the beginning of every generation, two strings are randomly generated. The first string is the template string which is simply a random string filled with four possible values (0, 1, 2, 3) at every given position. The second string is generated by performing approximately 33% to 66% random mutations of the template string. This includes insertions of a random value, deletion at a given position, and a point mutation of changing one value to another value at a given point in the string. This generation process hopefully produces two strings that are mostly similar but at the same time contain significant differences in their sequences.

Using a fixed string length size for the two similar strings that will be aligned, we evolved a best set of parameters. Table 1 shows the variables and their possible range. We then used a series of string lengths (10, 20, 30, 40, 50, 60, 70, 80, 90, 100) and attempted to evolve the best set of parameters for each of these parameters.

In calculating the output of the ACO given a set of parameters, we first showed empirically that the distribution of scores followed a normal distribution. This allowed us to easily throw out the data points that were outside $\overline{x} \pm \sigma$ and calculate a corrected mean ($\overline{x}_{\text{corr}}$) during the evolution of the parameters. This prevents random extreme values from affecting the mean and is a valid procedure after verifying that the output from the ACO is normally distributed. We also measured the total time to run all the trials of for each parameter ($t_{trials}$). Using these two variables we can assign a score to each parameter ($P_i$) [1]:

$$\text{Score}(P_i) = \frac{\overline{x}_{\text{corr}}(i)^3}{t_{trials}(i)} \qquad (1)$$

Using this scoring function, we were able to make a relative ranking of the current population of the highest scoring parameters. We then kept the top 1% of the population as "parents." The next generation was then generated using various parents and copy mutations (see Appendix to see the actual implementation of the genetic algorithm).

Several runs for each string length was preformed, in hopes of finding a set of parameters that came up often which would represent a locally optimal solution. After the best parameters for each of the strings were found, we can interpolate and approximate the best set of parameters for any string length up to a length of 100 characters.

# Results

## Ant Colony Optimization Results

Using the same string generators for the genetic algorithm, a string length of 50 was set. Then a random set of parameters were chosen. Afterwards, the ACO was run with these parameters and the chosen strings for 100 repetitions. The outputting scores were then gathered and compiled into a histogram. This process was repeated several times. Figure 1 shows one such histogram. All of the outputs from the ACOs followed a similar distribution to the one shown in Figure 1.

Hence we were able to conclude that to a first order approximation, we can expect the ACO output to be normally distributed. Given that the distribution is normal, we use this knowledge to save time during the genetic algorithm run and to make a good approximation of the mean score given by a set of parameters as was shown above.

## Genetic Algorithm Results

We explored the range of 10 - 100 characters for the string length. Our genetic algorithm probed this string length range by attempting to find the best set of parameters in 10 character increments: we ran a separate genetic algorithm for the string lengths 10, 20, 30, 40, 50, 60, 70, 80, 90, and 100 characters. Each of these string lengths were run multiple times with a population of 500 parameter sets, and the genetic algorithm was stopped after the population had reached an equilibrium for about 10 generations.

---

[1] It was found that if $\overline{x}_{\text{corr}}$ is not more emphasized than time, then the GA will start sacrificing score for faster and faster times. This is an undesirable result; raising to the third power provided better results.

Figure 2 and Figure 3 show the trends for each of the variables.[2] It is clear from these trends that most of the variables follow a linear trend. The only exceptions are $g$ and $a$ which both seem to require a lower limit and then proceed to grow starting at 80 string length. (There is not enough data to support what kind of growth the trend follows.) All other variables increase linearly, decrease linearly, or remain constant.

Using these variables we were able to interpolate values for any string length between 10 to 100. Using this system of approximation, we were able to design an optimized form of the ACO implementation that first calculated approximate optimal values for the parameter set and then runs the ACO on the input strings.

## Optimized ACO Results

Using the parameter set given above, we modified our ACO implementation to use the parameter table and interpolate approximately optimal parameters for a given input string (See Appendix A for the actual implementation). We provide here some of the sample alignments outputted by our implementation and compare them to the Needleman-Wunsch sequence alignment.

We divide the category of pairs of sequences into one of three types: highly similar (Type I), unalignable (Type II), and in between (Type III). This provides a qualitative basis for comparison between the Needleman-Wunsch and our ACO implementation.

To test highly similar pairs of sequences, we choose sequences that were nearly the same. Thus Type I alignments provided a sanity check to make sure that the our algorithm did indeed provide a satisfactory alignment.

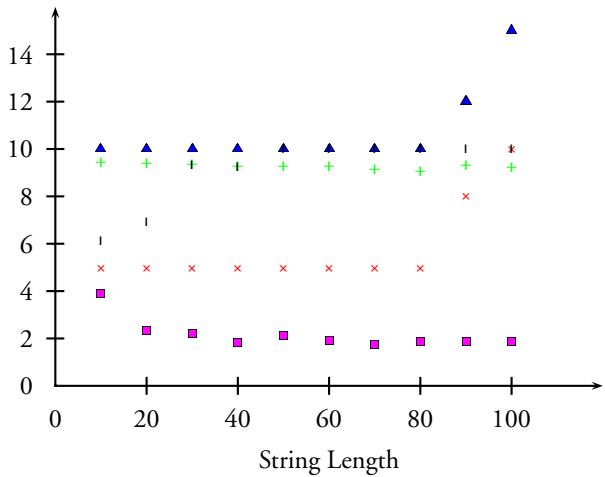

**Figure 2:** This figure shows the trends for the variables in the range of 0 to 15. ▲ represents $g$. × represents $a$. + represents $W_\phi$. ┆ represents $W_m$. ■ represents $W_r$.

In testing unalignable pairs of sequences, we choose sequences that were reversed of each other and no character appears twice in these sequences. Type II alignments provide a test to see how the ACO deals with pairs of sequences that it had not evolved to handle.

In order to test Type III, we generate a pair of sequences in the similar manner as we did during the evolutions of the parameters for genetic algorithms. Theoretically, it is these sequences that should provide the most interesting alignments.

Table 2 shows an example alignment for each one of these categories. There is not an objective measurement that would provide a quantitative degree of similarity between the alignments, hence we only provide a representative alignment of each of the categories instead.

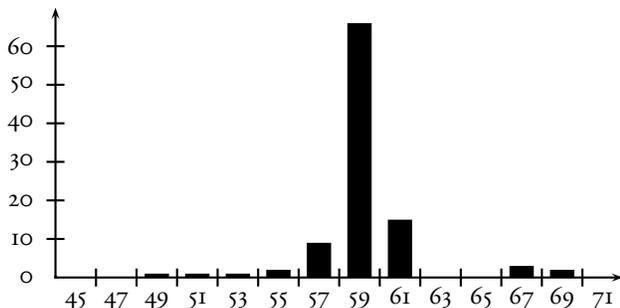

**Figure 1:** A histogram of output of one ACO run with same set of parameters for a 100 runs.

| Type | ACO | Needleman-Wunsch |
|---|---|---|
| I | ```abcdefgggghijklmnopq``` ``` |||||| |||||||||``` ```abcdefg---hijklmnopq``` | ```abcdefgggghijklmnopq``` ``` |||||| |||||||||``` ```abcdefg---hijklmnopq``` |
| II | ```qpo-nmlkjihgfedcba``` ``` |``` ```-abcdefghijklmnopq``` | ```qponmlkjihgfedcba``` ``` |``` ```abcdefghijklmnopq``` |
| III | ```-CACTTTTTCAGATCTATTG``` ``` ||||||||||||||| ||``` ```CTACTTTTTCAGATATATTC``` | ```C-ACTTTTTCAGATCTATTG``` ``` | ||||||||||||| ||``` ```CTACTTTTTCAGATATATTC``` |

**Table 2:** Sample alignments from both ACO and Needleman-Wunsch. Type I are pairs of sequences that are highly similar. Type II are pairs of sequences that are unalignable. Finally, Type III are pairs of sequences that are somewhere in between.

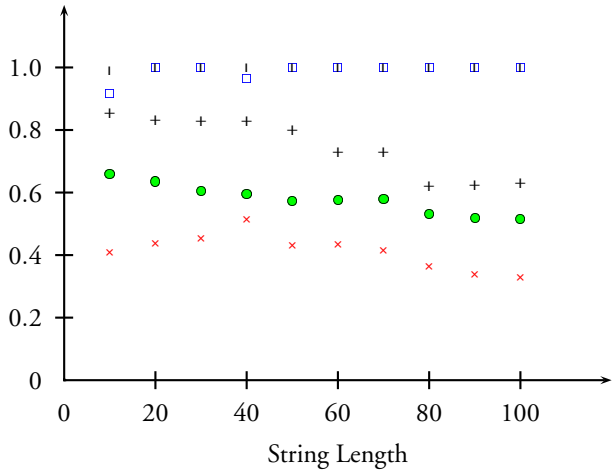

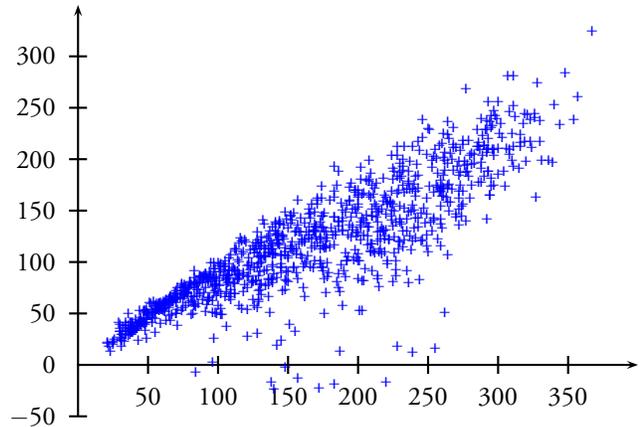

**Figure 3**: This figure shows the trends for the variables in the range of 0 to 1. + represents $\phi_0$. × represents $\phi_{step}$. ● represents $q_g$. ▫ represents $q_g$. ' represents $p$.

**Figure 4**: This is a scatter plot of scores from both Needleman-Wunsch and ACO. The x-axis is the Needleman-Wunsch scores, and the y-axis is the ACO scores.

## Discussion

### Analysis

At first glance, Table 2 provides some very promising results. It shows that the ACO actually preforms at a level of reaching nearly the mathematically optimal solution. In Type I, the alignments are exactly the same. For Type II, the ACO alignment is suboptimal compared with the Needleman-Wunsch because there is an extra gap that is not necessary. For Type III, the alignment is again nearly optimal: the main difference comes at the beginning of the alignment where the ACO could have made an extra match. It is interesting to note that on several other alignments, the main difference between ACO and Needleman-Wunsch comes at the beginning, a matter that we will address later.

Hence even from a completely qualitative standpoint, it can be said that the ACO does provide satisfactory results. While it can not be considered to be faster or more efficient than Needleman-Wunsch, it can, nevertheless, be applied to sequence alignment.

### Conclusion

As the results of the optimized ACO runs have shown, the ACO algorithm is a powerful algorithm that can be readily applied to sequence alignment. While it is true that this particular implementation of the ACO with respect to sequence alignment is not algorithmically faster or more efficient than the Needleman-Wunsch dynamic programming equivalent, the purpose of this paper was to show that the ACO is a viable algorithm that can be used to solve computational biology problems perhaps more efficiently when searching an NP hard search space.

Some problems and sources of error were encountered during the course of experimental runs of our implementation. We realize that there is an error in our genetic algorithm in that we should have optimized for high scoring consistency and low run time instead of optimizing for high *average score* and low run time. This produced parameter sets for the ACO that were suboptimal for the problem that we were addressing. Due to the time constraint of the project, we were not able to fix this issue in time for the deadline.

Also, the higher string lengths such as the 90 and 100 case were based on data that was not completely reliable in that we did not have enough time to run the genetic algorithm multiple times with many generations due to the highly computationally expensive nature of this algorithm. This produced possibly suboptimal parameter sets for these string lengths and hence cannot be as well trusted as the parameter sets generated for the lower part of the string length range.

Finally, we realize that genetic algorithms have the danger of producing suboptimal results, but we are confident in the results we have obtained because multiple runs of the genetic algorithm for a given string length gave results that were very similar to each other, an indication that the parameter sets generated were indeed near the global optimum.

A few improvements could be made to the ACO implementation for sequence alignment as well. The way that it is currently implemented, there is a limitation in the accuracy of the outputted alignments. The ants crawling up from the bottom right of the scoring matrix to the upper left cause alignments to be very accurate at the beginning of the traversal but not at the end of the alignment since the paths of the ants will diverge towards the end. This limitation can be easily overcome by adding an additional feature of having non-scoring ants dropped in random places on the scoring matrix. This way, areas around where the ants were dropped will have better alignment paths for when the scoring ants which started in the lower left corner reached the paths that were reinforced by the non-scoring ants. If local alignment is desired, then the random ants can be simply considered to be scoring.

In conclusion, we found that the ACO algorithm is an algorithm that can be used to solve biological problems. While in this case, it is not necessarily better than the existing algorithm for two sequence alignment; we are confident that the ACO can be readily adapted for much more complicated problems such as Multi-Sequence Alignment (MSA). In this case, the search space is much larger and the Needleman-Wunsch algorithm is out of the question; hence, a good adaption of the ACO meta-heuristic might provide a solution that rivals even the fastest MSA implementation today. Additional applications of ACO to biological problems could be such problems as clustering or even an ACO adapted algorithm for protein folding.

## Contributions

Thanks to Qian Zhang for his suggestions on ways of improving our algorithm. To Charles Duan for helping us with LaTeX. To the Harvard NICE computing environment. To our TF, Jon Radoff for answering our never-ending questions. And to Professor Church for his instruction of this class.

# Appendix

# A   ACO Implementation

```perl
#!/usr/local/bin/perl
use strict;
# Ant.pl
# Authors:  Li-Wei King, Aaron Lee

################################
# Default values just in case something goes terribly wrong
my $NUM_ANTS      = 10;
my $MAX_GEN       = 50;
my $PHER_STEP     = .1;
my $PHER_WEIGHT   = .1;
my $MATCH_WEIGHT  = .1;
my $REGION_WEIGHT = .00001;
my $PROB_PROB     = .5;           # Must be between 0 and 1.  It is the probability
                                  #   that the direction will be chosen based on probability.
my $INIT_PHER = .01;              # The amount of pheremone that an untraveled path has.
my $L_DECAY = .7;                 # Local decay.  Cannot be larger than 1.
my $G_DECAY = .7;                 # Global decay.  Cannot be larger than 1.

# Constants that won't eventually be passed to the program.
use constant MATCH_BONUS => 5;
use constant GAP_PENALTY => -4;
use constant MISMATCH_PENALTY =>  -3;

############
# Useful abbreviations
#For encoding the direction an ant moves
use constant D =>  0;
use constant L =>  1;
use constant U =>  2;
use constant N => -1;          # Undefined direction

# The index at which certain information is stored in the $ants array
use constant X_INDEX =>  0;
use constant Y_INDEX =>  1;
use constant DIR_INDEX =>  2;    # Where the ant decides to move from that position
use constant SCORE_INDEX =>  3;  # The current score at that position

############
# Table of GA optimized values
my @GA_TABLE;
$GA_TABLE[10][0] = 10; $GA_TABLE[20][0] = 10; $GA_TABLE[30][0] = 10;
$GA_TABLE[40][0] = 10; $GA_TABLE[50][0] = 10; $GA_TABLE[60][0] = 10;
$GA_TABLE[70][0] = 10; $GA_TABLE[80][0] = 10; $GA_TABLE[90][0] = 12;
$GA_TABLE[100][0] = 15;

$GA_TABLE[10][1] = 5; $GA_TABLE[20][1] = 5; $GA_TABLE[30][1] = 5;
$GA_TABLE[40][1] = 5; $GA_TABLE[50][1] = 5; $GA_TABLE[60][1] = 5;
$GA_TABLE[70][1] = 5; $GA_TABLE[80][1] = 5; $GA_TABLE[90][1] = 8;
$GA_TABLE[100][1] = 10;

$GA_TABLE[10][2] = .411191490;  $GA_TABLE[20][2] = .438349294;
$GA_TABLE[30][2] = .453581583;  $GA_TABLE[40][2] = .517059770;
$GA_TABLE[50][2] = .432201854;  $GA_TABLE[60][2] = .436950953;
$GA_TABLE[70][2] = .417636990;  $GA_TABLE[80][2] = .366514982;
$GA_TABLE[90][2] = .341437549;  $GA_TABLE[100][2] = .329430526;

$GA_TABLE[10][3] = 9.434392207; $GA_TABLE[20][3] = 9.423857194;
$GA_TABLE[30][3] = 9.353249096; $GA_TABLE[40][3] = 9.284172996;
$GA_TABLE[50][3] = 9.290221874; $GA_TABLE[60][3] = 9.282690356;
$GA_TABLE[70][3] = 9.149204714; $GA_TABLE[80][3] = 9.064648465;
$GA_TABLE[90][3] = 9.332940631; $GA_TABLE[100][3] = 9.259328124;

$GA_TABLE[10][4] = 6.109820365; $GA_TABLE[20][4] = 6.926580738;
$GA_TABLE[30][4] = 9.343075608; $GA_TABLE[40][4] = 9.244311660;
$GA_TABLE[50][4] = 10;          $GA_TABLE[60][4] = 10;
$GA_TABLE[70][4] = 10;          $GA_TABLE[80][4] = 10;
$GA_TABLE[90][4] = 10;          $GA_TABLE[100][4] = 10;

$GA_TABLE[10][5] = 3.909960135; $GA_TABLE[20][5] = 2.350289525;
$GA_TABLE[30][5] = 2.224402772; $GA_TABLE[40][5] = 1.853908945;
$GA_TABLE[50][5] = 2.142958734; $GA_TABLE[60][5] = 1.915834968;
$GA_TABLE[70][5] = 1.736982155; $GA_TABLE[80][5] = 1.878705913;
```

```perl
 74   $GA_TABLE[90][5]  = 1.856936304; $GA_TABLE[100][5] = 1.862138526;
 75
 76   $GA_TABLE[10][6]  = 0.853763237; $GA_TABLE[20][6]  = 0.830586273;
 77   $GA_TABLE[30][6]  = 0.827315180; $GA_TABLE[40][6]  = 0.827352487;
 78   $GA_TABLE[50][6]  = 0.798933405; $GA_TABLE[60][6]  = 0.728488635;
 79   $GA_TABLE[70][6]  = 0.730894471; $GA_TABLE[80][6]  = 0.620157623;
 80   $GA_TABLE[90][6]  = 0.622498305; $GA_TABLE[100][6] = 0.628942392;
 81
 82   $GA_TABLE[10][7]  = 0.660078498; $GA_TABLE[20][7]  = 0.635274652;
 83   $GA_TABLE[30][7]  = 0.606102419; $GA_TABLE[40][7]  = 0.595236898;
 84   $GA_TABLE[50][7]  = 0.571805707; $GA_TABLE[60][7]  = 0.577967010;
 85   $GA_TABLE[70][7]  = 0.579124731; $GA_TABLE[80][7]  = 0.532124853;
 86   $GA_TABLE[90][7]  = 0.519128179; $GA_TABLE[100][7] = 0.515925041;
 87
 88   $GA_TABLE[10][8]  = 0.917907684; $GA_TABLE[20][8]  = 1;
 89   $GA_TABLE[30][8]  = 1;           $GA_TABLE[40][8]  = 0.965550166;
 90   $GA_TABLE[50][8]  = 1;           $GA_TABLE[60][8]  = 1;
 91   $GA_TABLE[70][8]  = 1;           $GA_TABLE[80][8]  = 1;
 92   $GA_TABLE[90][8]  = 1;           $GA_TABLE[100][8] = 1;
 93
 94   $GA_TABLE[10][9]  = 0.990544051; $GA_TABLE[20][9]  = 1;
 95   $GA_TABLE[30][9]  = 1;           $GA_TABLE[40][9]  = 1;
 96   $GA_TABLE[50][9]  = 1;           $GA_TABLE[60][9]  = 1;
 97   $GA_TABLE[70][9]  = 1;           $GA_TABLE[80][9]  = 1;
 98   $GA_TABLE[90][9]  = 1;           $GA_TABLE[100][9] = 1;
 99
100
101   ######################################
102   if ($#ARGV != 1) {
103       print "Usage: \n";
104       print "    ant.pl <SEG1> <SEG2>\n";
105       exit;
106   }
107
108   my $SEG1 = $ARGV[0];
109   my $SEG2 = $ARGV[1];
110
111   my $strlenavg = length($SEG1) + length($SEG2) / 2;
112   if ($strlenavg < 10 || $strlenavg > 100) {
113       print "The string lengths of SEG1 and SEG2 must have an average string
114   length betweeen 10 and 100!\n";
115       exit;
116   }
117
118   optimize_vars_run_aco($SEG1, $SEG2, $strlenavg);
119
120   sub optimize_vars_run_aco {
121       my $str1 = shift;
122       my $str2 = shift;
123       my $len = shift;
124       my $opt;
125       my $lowten = (int $len / 10) * 10;
126       my $highten = (int $len / 10) * 10 + 10;
127       my $wherex = $len - $lowten;
128
129       for (my $i = 0; $i <= 9; $i++) {
130           push(@opt, $wherex * ($GA_TABLE[$highten][$i]-$GA_TABLE[$lowten][$i])/10 + $GA_TABLE[$lowten][$i]);
131       }
132
133       run_aco($str1, $str2, $opt[0], $opt[1], $opt[2], $opt[3], $opt[4], $opt[5],
134       $opt[6], $opt[7], $opt[8], $opt[9]);
135   }
136
137   sub run_aco {
138       # Arguments to the array
139       my $seq1 = shift;
140       my $seq2 = shift;
141       my $MAX_GEN = shift;
142       my $NUM_ANTS = shift;
143       my $PHER_STEP = shift;
144       my $PHER_WEIGHT = shift;
145       my $MATCH_WEIGHT = shift;
146       my $REGION_WEIGHT = shift;
147       my $INIT_PHER = shift;
148       my $L_DECAY = shift;
149       my $G_DECAY = shift;
150       my $prob_prob = shift;
151
152       #Other variables
153       my $high_score;
```

```perl
my @best_path_x;
my @best_path_y;
my @best_path_dir;
my $prev_score = 0;   # Used to figure out if there are repeats or not
my $repeat_counter = 0;

print "Aligning: $seq1 with $seq2\n";

# Store the sequences in arrays
my @seqy = split('', $seq1);
my @seqx = split('', $seq2);

my @pher;             # Stores the pheromone values
# begin matrix initialization.
for (my $i = 0; $i <= $#seqy; $i++) {
    for (my $j = 0; $j <= $#seqx; $j++) {
        $pher[$i][$j][D] = $INIT_PHER;
        $pher[$i][$j][L] = $INIT_PHER;
        $pher[$i][$j][U] = $INIT_PHER;
    }
}

# begin main generation loop
for (my $gen = 0; $gen < $MAX_GEN; $gen++) {
    my %done = ();    # This hash stores a 1 if the ant is finished, 0 if it isn't
        my @ants = ();    # Stores the information for all the ants

        # reinitialize ants.  The ant array is organized thus:
        # $ants[ant_id][move_number][information for that move]
        for (my $ant = 0; $ant < $NUM_ANTS; $ant++) {
            $ants[$ant][0][X_INDEX] = $#seqx;
            $ants[$ant][0][Y_INDEX] = $#seqy;
            $done[$ant] = 0;
            if ($seqx[$#seqx] eq $seqy[$#seqy]) {
                #$ants[$ant][0][SCORE_INDEX] = MATCH_BONUS;
                $ants[$ant][0][SCORE_INDEX] = 0;
            } else {
                $ants[$ant][0][SCORE_INDEX] = 0;
            }
        }

    my $ants_done = 0;
    for (my $step = 0; $ants_done < $NUM_ANTS; $step++) {

        # simulate the ants:
        for (my $ant = 0; $ant < $NUM_ANTS; $ant++) {

            #If the ant is done, then skip it.
            if ($done[$ant] == 1) {
                next;
            }

            # pick up where we left off:
            my $ant_x = $ants[$ant][$step][X_INDEX];
            my $ant_y = $ants[$ant][$step][Y_INDEX];

            # get the direction that the ant should take, and update
            # the direction index
            my $dir = get_dir($ant_x, $ant_y,\@pher, $PHER_WEIGHT, $MATCH_WEIGHT,
                $REGION_WEIGHT, $prob_prob, \@seqx, \@seqy);
            $ants[$ant][$step][DIR_INDEX] = $dir;

            # update pher matrix
            $pher[$ant_y][$ant_x][$dir] += $PHER_STEP;
            # Do local decay
            $pher[$ant_y][$ant_x][$dir] *= $L_DECAY;

            # move in the chosen direction and update the score.
            # The score stored at each square includes
            # match/mismatch score for this square as well as well
            # as any gap penalty for the move that got the ant to
            # that position .
            if ($dir == D) {
                $ants[$ant][$step + 1][X_INDEX] = $ant_x - 1;
                $ants[$ant][$step + 1][Y_INDEX] = $ant_y - 1;
                $ants[$ant][$step + 1][SCORE_INDEX] = $ants[$ant][$step][SCORE_INDEX];
                if ($seqx[$ants[$ant][$step][X_INDEX]] eq $seqy[$ants[$ant][$step][Y_INDEX]]) {
                    $ants[$ant][$step + 1][SCORE_INDEX] += MATCH_BONUS;
                } else {
```

```perl
234                    $ants[$ant][$step + 1][SCORE_INDEX] += MISMATCH_PENALTY;
235                }
236
237            } elsif ($dir == U) {
238                $ants[$ant][$step + 1][X_INDEX] = $ant_x;
239                $ants[$ant][$step + 1][Y_INDEX] = $ant_y - 1;
240                $ants[$ant][$step + 1][SCORE_INDEX] = $ants[$ant][$step][SCORE_INDEX] + GAP_PENALTY;
241            } elsif ($dir == L) {
242                $ants[$ant][$step + 1][X_INDEX] = $ant_x - 1;
243                $ants[$ant][$step + 1][Y_INDEX] = $ant_y;
244                $ants[$ant][$step + 1][SCORE_INDEX] = $ants[$ant][$step][SCORE_INDEX] + GAP_PENALTY;
245            }
246
247            #Check if the ant is done and update score if needed
248            if ($ants[$ant][$step + 1][X_INDEX] == 0 || $ants[$ant][$step + 1][Y_INDEX] == 0) {
249                $ants[$ant][$step + 1][DIR_INDEX] = N;
250                $done{$ant} = 1;
251                if ($seqx[$ants[$ant][$step+1][X_INDEX]] eq $seqy[$ants[$ant][$step+1][Y_INDEX]]) {
252                    $ants[$ant][$step + 1][SCORE_INDEX] += MATCH_BONUS;
253                } else {
254                    $ants[$ant][$step + 1][SCORE_INDEX] += MISMATCH_PENALTY;
255                }
256
257                $ants_done++;
258            }
259        }
260    }
261
262    #Do a global update
263    #Find the best score
264    my $max_gen_score = $ants[0][$#{$ants[0]}][SCORE_INDEX];
265    my $max_score_ant = 0;
266    my @best_gen_path_x = ();      # The best path of the generation:  x coordinates
267    my @best_gen_path_y = ();
268    my @best_gen_path_dir = ();
269
270    for (my $a = 1; $a <= $#ants; $a++) {
271        if ($ants[$a][$#{$ants[$a]}][SCORE_INDEX] > $max_gen_score) {
272            $max_gen_score = $ants[$a][$#{$ants[$a]}][SCORE_INDEX];
273            $max_score_ant = $a;
274        }
275    }
276
277    #Get the best path
278    for(my $s = 0; $s <= $#{$ants[$max_score_ant]}; $s++) {
279        push (@best_gen_path_x, $ants[$max_score_ant][$s][X_INDEX]);
280        push (@best_gen_path_y, $ants[$max_score_ant][$s][Y_INDEX]);
281        push (@best_gen_path_dir, $ants[$max_score_ant][$s][DIR_INDEX]);
282    }
283
284    #Find out if the high score this time is the all time high.
285    if ($gen == 0) {
286        $high_score = $max_gen_score;
287        for (my $i = 0; $i <= $#best_gen_path_x; $i++) {
288            push (@best_path_x, $best_gen_path_x[$i]);
289            push (@best_path_y, $best_gen_path_y[$i]);
290            push (@best_path_dir, $best_gen_path_dir[$i]);
291        }
292    } elsif ($max_gen_score > $high_score) {
293        @best_path_x = ();
294        @best_path_y = ();
295        @best_path_dir = ();
296        $high_score = $max_gen_score;
297        for (my $i = 0; $i <= $#best_gen_path_x; $i++) {
298            push (@best_path_x, $best_gen_path_x[$i]);
299            push (@best_path_y, $best_gen_path_y[$i]);
300            push (@best_path_dir, $best_gen_path_dir[$i]);
301        }
302    }
303
304    #Reinforce the best path again based on how good it is.  I'm
305    #using < rather than <= because the last direction is undefined.
306    if ($max_gen_score > 0 && $high_score > 0) {
307        for (my $i = 0; $i < $#best_gen_path_x; $i++) {
308            $pher[$best_gen_path_y[$i]][$best_gen_path_x[$i]][$best_gen_path_dir[$i]]
309                += ($max_gen_score/$high_score) * $PHER_STEP;
310        }
311    }
312
313    #global_decay(\@pher, $G_DECAY);
```

```perl
314
315          #Check for convergence
316          if ($max_gen_score == $prev_score) {
317              $repeat_counter++;
318          } else {
319              $max_gen_score = $prev_score;
320              $repeat_counter = 0;
321          }
322
323          #If we get the same score 5 times, we'll stop the calculation
324          if ($repeat_counter == 5) {
325              last;
326          }
327      }
328
329      my @alignment = get_alignment(\@best_path_x, \@best_path_y, \@seqx, \@seqy);
330      my $x_alignment = $alignment[0];
331      my $dashes = $alignment[1];
332      my $y_alignment = $alignment[2];
333
334      print "The best alignment obtained was: \n";
335      print "$x_alignment\n";
336      print "$dashes\n";
337      print "$y_alignment\n";
338      print "The high score was: $high_score \n";
339
340  }
341
342  sub get_alignment {
343      #Arguments
344      my $best_x_ref = shift;   # Stores the x coordinates of the best path
345      my $best_y_ref = shift;
346      my $seq_x_ref = shift;
347      my $seq_y_ref = shift;
348
349      #Other variables
350      my $last_x;               # The x coordinate of the last step we traced
351      my $last_y;
352      my $x_result = "";
353      my $y_result = "";
354      my $dashes = "";
355
356
357      # Handle the cases where the alignment doesn't start in the 0,0
358      # position
359      if($best_x_ref->[$#{@$best_x_ref}] != 0) {
360          for (my $i = 0; $i < $best_x_ref->[$#{@$best_x_ref}]; $i++) {
361              $x_result = $x_result.$seq_x_ref->[$i];
362              $y_result = $y_result."-";
363              $dashes = $dashes." ";
364          }
365      } elsif ($best_y_ref->[$#{@$best_y_ref}] != 0) {
366          for (my $i = 0; $i < $best_y_ref->[$#{@$best_y_ref}]; $i++) {
367              $x_result = $x_result."-";
368              $y_result = $y_result.$seq_y_ref->[$i];
369              $dashes = $dashes." ";
370          }
371      }
372
373      #First step.  Start from the beginning of the alignment (the end of
374      # the best_x_ref and best_y_ref arrays
375      $last_x = $best_x_ref->[$#{@$best_x_ref}];
376      $last_y = $best_y_ref->[$#{@$best_y_ref}];
377      $x_result = $x_result.$seq_x_ref->[$last_x];
378      $y_result = $y_result.$seq_y_ref->[$last_y];
379
380      if ($seq_x_ref->[$last_x] eq $seq_y_ref->[$last_y]) {
381          $dashes = $dashes."|";
382      } else {
383          $dashes = $dashes." ";
384      }
385
386      for (my $i = $#{@$best_x_ref} - 1; $i >= 0; $i--) {
387          # Gap in the x if the x index has not changed.  Append a gap to
388          # the sequence.
389          if ($best_x_ref->[$i] eq $last_x) {
390              $last_y = $best_y_ref->[$i];
391              $x_result = $x_result."-";
392              $y_result = $y_result."$seq_y_ref->[$best_y_ref->[$i]]";
393              $dashes = $dashes." ";
```

```perl
394          } elsif ($best_y_ref->[$i] eq $last_y) {
395              # Gap in y
396              $last_x = $best_x_ref->[$i];
397              $y_result = $y_result."-";
398              $x_result = $x_result."$seq_x_ref->[$best_x_ref->[$i]]";
399              $dashes = $dashes." ";
400          } else {
401              #Diagonal move
402              $last_x = $best_x_ref->[$i];
403              $last_y = $best_y_ref->[$i];
404              $y_result = $y_result."$seq_y_ref->[$best_y_ref->[$i]]";
405              $x_result = $x_result."$seq_x_ref->[$best_x_ref->[$i]]";
406              if ($seq_y_ref->[$best_y_ref->[$i]] eq $seq_x_ref->[$best_x_ref->[$i]]) {
407                  $dashes = $dashes."|";
408              } else {
409                  $dashes = $dashes." ";
410              }
411          }
412      }

414      return ($x_result, $dashes, $y_result);
415  }

417  sub global_decay {
418      #Arguments
419      my $pher_ref = shift;
420      my $g_decay = shift;

422      for (my $x = 0; $x <= $#{ $pher_ref }; $x++) {
423          for (my $y = 0; $y <= $#{ $pher_ref->[$x] }; $y++ ) {
424              for (my $d = 0; $d < 3; $d++) {
425                  $pher_ref->[$x]->[$y]->[$d] *= $g_decay;
426              }
427          }
428      }
429  }

430  sub get_dir {
432      #Arguments
433      my $x = shift;
434      my $y = shift;
435      my $pher_ref = shift;
436      my $pher_weight = shift;
437      my $match_weight = shift;
438      my $region_weight = shift;
439      my $prob_prob = shift;
440      my $seqx = shift;
441      my $seqy = shift;

443      #Calculate the regional weighting for these coordinates.  This
444      #assumes that both sequences are the same length.
445      #TODO: MAKE SURE TO CHANGE REGIONAL WEIGHTINGS

447      my $reg_w_up;
448      my $reg_w_left;
449      my $reg_w_diag;

451      if ($x == $y) {
452          $reg_w_up = 1;
453          $reg_w_left = 1;
454          $reg_w_diag = 2;
455      } elsif ($x > $y) {
456          $reg_w_up = 1;
457          $reg_w_diag = 1.5;
458          $reg_w_left = 2;
459      } else {
460          $reg_w_up = 2;
461          $reg_w_left = 1;
462          $reg_w_diag = 1.5;
463      }

465      #Figure out whether going up, diagonal, or left will give us a match
466      my $match_up;
467      my $match_left;
468      my $match_diag;

470      if ($seqx->[$x] eq $seqy->[$y - 1]) {
471          $match_up = 2;
472      } else {
473          $match_up = 1;
```

```perl
474          }

476          if ($seqx->[$x - 1] eq $seqy->[$y - 1]) {
477              $match_diag = 2;
478          } else {
479              $match_diag = 1;
480          }

482          if ($seqx->[$x - 1] eq $seqy->[$y]) {
483              $match_left = 2;
484          } else {
485              $match_left = 1;
486          }

488          #Figure out the scores for each direction
489          my $up_score;
490          my $diag_score;
491          my $left_score;

493          $up_score = exp(log($pher_ref->[$y]->[$x]->[U]) * $pher_weight
494                  + log($match_up) * $match_weight
495                  + log($reg_w_up) * $region_weight);

497          $diag_score = exp(log($pher_ref->[$y]->[$x]->[D]) * $pher_weight
498                  + log($match_diag) * $match_weight
499                  + log($reg_w_diag) * $region_weight);

501          $left_score = exp(log($pher_ref->[$y]->[$x]->[L]) * $pher_weight
502                  + log($match_left) * $match_weight
503                  + log($reg_w_left) * $region_weight);

505          #Decide whether we're gonna decide the direction based on the best
506          #score, or based on a weighted probability.
507          #
508          if (rand(1) > $prob_prob) {
509              #If we're just gonna do it based on best score

511              my $high_score = max($up_score, $diag_score, $left_score);

513              if ($high_score == $up_score) {
514                  return U;
515              } elsif ($high_score == $diag_score) {
516                  return D;
517              } else {
518                  return L;
519              }
520          } else {
521              #Determine the direction using probabilities

523              my $total_score = $up_score + $diag_score + $left_score;

525              my $rand_num = rand($total_score);

527              $total_score -= $up_score;
528              if ($rand_num > $total_score) {
529                  return U;
530              }

532              $total_score -= $diag_score;
533              if ($rand_num > $total_score) {
534                  return D;
535              } else {
536                  return L;
537              }
538          }
539  }

541  sub max {

543      my $max = $_[0];

545      for(my $i = 1; $i <= $#_; $i++) {
546          if ($_[$i] > $max) {
547              $max = $_[$i];
548          }
549      }

551      return $max;
552  }
```

# B  Genetic Algorithm Implementation

```perl
1    #!/usr/local/bin/perl -w
2    use Time::HiRes;
3    use strict;
4
5    ### BEGIN VARIABLES
6
7    # Constants
8    my $STR_LENGTH       = 20;
9    my $POP_LIMIT        = 1000;
10   my $RECORD_FREQ      = 5;
11   my $NUM_TEST_IND     = 7;
12   my $KEEP_PARENTS     = int ($POP_LIMIT * .1);
13   my $CROSSOVER        = .2;
14   my $ZERO             = 0.0000000001;
15
16   # Global variables
17   my @pop;
18   my $cur_gen          = 0;
19   my $file_num         = 0;
20   my $SCORE_INDEX = 10;
21
22   my @range = ();
23
24   $range[0][0] = 10;              # $MAX_GEN
25   $range[0][1] = 40;
26   $range[1][0] = 5;              # $NUM_ANTS
27   $range[1][1] = 30;
28   $range[2][0] = $ZERO;          # $PHER_STEP
29   $range[2][1] = 1;
30   $range[3][0] = $ZERO;          # $PHER_WEIGHT
31   $range[3][1] = 10;
32   $range[4][0] = $ZERO;          # $MATCH_WEIGHT
33   $range[4][1] = 10;
34   $range[5][0] = $ZERO;          # $REGION_WEIGHT
35   $range[5][1] = 5;
36   $range[6][0] = $ZERO;          # $INIT_PHER
37   $range[6][1] = 1;
38   $range[7][0] = $ZERO;          # $L_DECAY
39   $range[7][1] = 1;
40   $range[8][0] = $ZERO;          # $G_DECAY
41   $range[8][1] = 1;
42   $range[9][0] = $ZERO;          # $PROB_PROB
43   $range[9][1] = 1;
44
45
46   ### BEGIN MAIN PROGRAM
47
48   #randomize variables
49   randomize_vars();
50
51   while (1) {
52       #initialization
53       my $seq1 = getrandseq();
54       my $seq2 = mutate($seq1);
55
56       print "Generation $cur_gen\n";
57       print "-Testing sequences:\n";
58       print "-Sequence 1 (".length($seq1)."): $seq1\n";
59       print "-Sequence 2 (".length($seq2)."): $seq2\n";
60
61
62       test_pop($seq1, $seq2);
63
64       gen_next_pop();
65
66       if ($cur_gen == $RECORD_FREQ) {
67           print_pop();
68           $cur_gen = 0;
69           $file_num++;
70       }
71
72       $cur_gen++;
73   }
74
75   ### BEGIN SUBROUTINES
76
77   sub getrandseq {
```

```perl
        my $limit = $STR_LENGTH;
        my $str = "";

        for (my $i = 0; $i < $limit; $i++) {
            my $aa = int rand(4);

            $str .= $aa;
        }

        return $str;
}

sub mutate {
        my $template = shift;
        my $num_muts = int length($template) / (rand (2) + 1.5);

        for (my $i = 0; $i < $num_muts; $i++) {
            my $whichmut = int rand(3);
            my $where = int rand(length($template));

            if ($whichmut == 0) {
                # Point mutation
                substr($template,$where,1,int rand(4));
            } elsif ($whichmut == 1) {
                # Insertion
                my $beg = substr($template,0,$where);
                my $end = substr($template,$where);
                my $num = int rand(4);
                $template = "$beg$num$end";
            } else {
                # Deletion
                my $beg = substr($template,0,$where);
                my $end = substr($template,$where+1);
                $template = "$beg$end";
            }
        }

        return $template;
}

sub randomize_vars {

        for (my $i = 0; $i < $POP_LIMIT; $i++) {
            for (my $j = 0; $j <= $#range; $j++) {
                $pop[$i][$j] = rand($range[$j][1]-$range[$j][0]) + $range[$j][0];
            }
            $pop[$i][0] = int $pop[$i][0];
            $pop[$i][1] = int $pop[$i][1];
        }
}

sub test_pop {
        my $seq1 = shift;
        my $seq2 = shift;

        my @winners;

        print "-- Beginning population testing...\n";
        for (my $i = 0; $i < $POP_LIMIT; $i++) {
            my $score = score_individual($i, $seq1, $seq2);
            $pop[$i][$SCORE_INDEX] = $score;
        }

        # Sort so that the winners are at the top and losers are at the bottom
        @pop = sort {$b->[$SCORE_INDEX] <=> $a->[$SCORE_INDEX]} @pop;

        print "-- DONE! \n";
        print "-- Top three scores: ";
        print "$pop[0][$SCORE_INDEX], $pop[1][$SCORE_INDEX], $pop[2][$SCORE_INDEX]\n";
}

sub score_individual {
        my $ind = shift;
        my $seq1 = shift;
        my $seq2 = shift;

        my @scores = ();

        my $totalscore = 0;
        my $totaltime = 0;
```

```perl
        print "-- Testing string length: $STR_LENGTH\n";
        print "-- Generation: ".($file_num*$RECORD_FREQ+$cur_gen)."\n";
        print "-- Testing individual $ind \n";
        print "----- Ants : $pop[$ind][1] \n";
        print "----- Gens : $pop[$ind][0] \n";
        for (my $i = 0; $i < $NUM_TEST_IND; $i++) {
            my $score;
            my $before = Time::HiRes::time();
            $score = run_aco($seq1, $seq2, int $pop[$ind][0], int $pop[$ind][1],
                                $pop[$ind][2], $pop[$ind][3], $pop[$ind][4],
                                $pop[$ind][5], $pop[$ind][6], $pop[$ind][7],
                                $pop[$ind][8], $pop[$ind][9]);
            $totaltime += Time::HiRes::time() - $before;
            $totalscore += $score;
            push(@scores, $score);
        }

        my $mean = $totalscore / $NUM_TEST_IND;
        print "------- Mean:$mean\n";

        my $stddev = 0;
        for (my $i = 0; $i < $NUM_TEST_IND; $i++) {
            my $term = ($mean - $scores[$i])*($mean - $scores[$i]);
            $stddev += $term;
        }
        $stddev = sqrt($stddev / ($NUM_TEST_IND + 1));
        print "------- Standard dev:$stddev\n";

        my $newmean = 0;
        my $datapoints = 0;
        my $thrownout = 0;
        for (my $i = 0; $i < $NUM_TEST_IND; $i++) {
            if ($scores[$i] >= $mean - $stddev && $scores[$i] <= $mean + $stddev) {
                $newmean += $scores[$i];
                $datapoints++;
            } else {
                $thrownout++;
            }
        }
        if ($datapoints == 0) {
            $newmean = -10000;
        } else {
            $newmean /= $datapoints;
        }

        my $ga_score = ($newmean * $newmean * $newmean) / $totaltime;
        print "------- Good datapoints: $datapoints\n";
        print "------- Thrown out data: $thrownout\n";
        print "------- Corrected mean: $newmean\n";
        print "------- Average time: ".$totaltime/$NUM_TEST_IND."\n";
        print "------- Final GA score: ".$ga_score."\n";

        return ($ga_score);
    }

    sub gen_next_pop {
        # Populate the next generation
        for (my $i = $KEEP_PARENTS; $i < $POP_LIMIT; $i++) {
            my $cur_parent = int rand($KEEP_PARENTS);

            for (my $j = 0; $j < $SCORE_INDEX; $j++) {
                $pop[$i][$j] = $pop[$cur_parent][$j];

                # Copy mutation
                my $which_mut = rand(10);
                if ($which_mut < 2) {
                    # Plus case
                    $pop[$i][$j] += .1 * rand(1) * ($range[$j][1] - $range[$j][0]);

                    if ($pop[$i][$j] > $range[$j][1]) {
                        $pop[$i][$j] = $range[$j][1];
                    }
                } elsif ($which_mut < 4) {
                    # Minus case
                    $pop[$i][$j] -= .1 * rand(1) * ($range[$j][1] - $range[$j][0]);

                    if ($pop[$i][$j] < $range[$j][0]) {
                        $pop[$i][$j] = $range[$j][0];
                    }
                }
```

```perl
238            } else {
239                # Do nothing case
240            }
241
242            # Branch migration
243            if (rand(1) < $CROSSOVER) {
244                $cur_parent = int rand($KEEP_PARENTS);
245            }
246        }
247    }
248 }
249
250 sub print_pop {
251     my $filename = "out-$file_num.txt";
252
253     open(FH, "> $filename");
254
255     printf FH ("%13s ", "MAX_GEN");
256     printf FH ("%13s ", "NUM_ANTS");
257     printf FH ("%13s ", "PHER_STEP");
258     printf FH ("%13s ", "PHER_WEIGHT");
259     printf FH ("%13s ", "MATCH_WEIGHT");
260     printf FH ("%13s ", "REGION_WEIGHT");
261     printf FH ("%13s ", "INIT_PHER");
262     printf FH ("%13s ", "L_DECAY");
263     printf FH ("%13s ", "G_DECAY");
264     printf FH ("%13s ", "PROB_PROB");
265     printf FH ("%13s ", "SCORE");
266     print FH "\n";
267
268
269     for (my $i = 0; $i < $POP_LIMIT; $i++) {
270         for (my $j = 0; $j <= $SCORE_INDEX; $j++){
271             printf FH ("%13.9f ", $pop[$i][$j]);
272         }
273         print FH "\n";
274     }
275     print FH "\n";
276     close FH;
277 }
278
279 ####################################
280 #  Put ACO code here
281 #
282 #    In the interest of space, the ACO code was not included here.
283 #        Please see Appendix A to see the implementation of the ACO.
284 #        The only change that was made was silencing all the output functions.
```